\documentclass[aps,12pt,superscriptaddress,amsfonts,amssymb,amsmath]{revtex4-2}
\pdfoutput=1

\usepackage{graphicx}
\usepackage{epsfig}
\usepackage{makeidx}
\usepackage{xcolor}
\usepackage{amsmath}
\usepackage{bm}
\usepackage{amssymb}

\begin{document}

\title{
Fluctuations in Aharonov-Bohm Electrodynamics} 

\author{F.\ O.\ Minotti \footnote{Email address: minotti@df.uba.ar}}
\affiliation{Universidad de Buenos Aires, Facultad de Ciencias Exactas y Naturales, Departamento de F\'{\i}sica, Buenos Aires, Argentina}
\affiliation{CONICET-Universidad de Buenos Aires, Instituto de F\'{\i}sica Interdisciplinaria y Aplicada (INFINA), Buenos Aires, Argentina}

\author{G.\ Modanese \footnote{Email address: giovanni.modanese@unibz.it}}
\affiliation{Free University of Bozen-Bolzano \\ Faculty of Engineering \\ I-39100 Bolzano, Italy}
\date{\today}

\linespread{0.9}

\begin{abstract}

We consider the application of the Fluctuation Dissipation Theorem (FDT) to the electrodynamics of Aharonov-Bohm (ABE), which differs from Maxwell's in that it allows for local non-conservation of charge. For the case of a system of non-conserved charges at thermal equilibrium we obtain the same spectral distribution of energy of the electromagnetic field as in Maxwell electrodynamics. However, the electric field contribution to that energy doubles that in Maxwell case, while the magnetic contribution is the same as in Maxwell theory, the electric excess energy is compensated by a negative contribution arising form the Aharonov-Bohm (AB) scalar field. For a conductor with local non-conservation of charge described by the $\gamma$ model, we derive the spectrum of current correlation at first order in $\gamma$, which results in a violet noise contribution added to the classical Johnson-Nyquist white noise result for the voltage fluctuations in a conductor.

\end{abstract}

\maketitle

\section{Introduction}
\label{intro}

\color{black}
Local conservation of charge (LCC) is expressed by the relation $\partial\rho/\partial t+\nabla\cdot\mathbf{j}=0$, where $\rho$ and $\mathbf{j}$ are the charge and current densities, respectively, the material sources of the electromagnetic field. It seems that LCC can hardly be disputed, particularly due to its direct relation to the well established gauge invariance of electromagnetic phenomena, both at the classical and quantum levels. Nevertheless, Aharonov and Bohm \cite{aharonov1963further} argued that LCC could be invalid at some small enough scale, which would require a description of electromagnetic phenomena in terms of the electromagnetic potentials, rather than the electric and magnetic fields. The theory they presented, which extends Maxwell electromagnetism to allow the inclusion of sources that do not satisfy strict LCC, is henceforth denoted as Aharonov-Bohm electrodynamics (ABE). In their original proposal Aharonov and Bohm considered that the scales at which fundamental laws, like LCC, are expected to fail were of the order of Planck's. However, first-principle calculations of electron currents in molecular devices, based on single-particle non-equilibrium Green's functions and density-functional theory, \cite{li2008definition,zhang2011first,walz2015local,cabra2018simulation} consistently indicated the violation of LCC at the molecular level. These are theoretical and numerical studies, so one can naturally argue that limitations in the modeling of the highly complex condensed-matter system could be responsible for the failure of LCC in those calculations. It is worth mentioning, however, that the failure of LCC was not substantially alleviated with the enlargement of the basis of states used to describe the Green's functions in the ab initio calculations. 
In line with this, in \cite{minotti_modanese_math13050892} LCC in many-body quantum mechanics was considered, using the formalism of non-equilibrium Green's functions by Baym and Kadanoff. In this formalism LCC can be expressed as a relation between one-particle and two-particle Green's functions (as well as in an alternative expression using the self-energy), that requires a limiting procedure in which the (imaginary) time arguments of two-particle Green's functions tend to each other. The argument used in \cite{minotti_modanese_math13050892} is that the time limiting procedure involves the high-frequency components of the Green's functions, and could thus be sensitive to the high-energy cutoff required to regularize the integrals in Feynman graphs corresponding to the Green's functions, particularly since it is not at present proved that this cutoff can be removed in the case of many-particle fermionic systems in three spatial dimensions \cite{salmhofer2019renormalization}. In particular, using a quasi-static approximation to the self-energy, it was also shown in \cite{minotti_modanese_math13050892} that a failure of LCC appears, quantitatively consistent with the first-principle calculations in molecular devices in \cite{zhang2011first}.

\color{black}

In this scenario, if applicable to a fermionic condensed-matter system such as a conductor or semiconductor, the apparent failure of LCC occurs at the level of the effective theory describing the low-energy, low-frequency dynamics. Strict local conservation is recovered only when the unresolved high-frequency current contributions associated with degrees of freedom not explicitly retained in the effective description are included in the local charge balance. These contributions may arise, for instance, from high-energy electron-hole excitations.

\color{black}
Within the reduced description, the effective non-conserved current may therefore give rise to the characteristic ABE phenomenology, including longitudinal electric fields and waves of pure potentials (gauge waves), which can interact with non-conserved effective currents \cite{minotti2023aharonov}. Here ABE is understood as an effective description obtained after the unresolved microscopic degrees of freedom have been eliminated. The current contributions associated with those degrees of freedom should not be regarded as additional independent sources to be simply superposed on the effective ABE current. If all microscopic degrees of freedom were explicitly restored, strict local charge conservation would also be restored, and the description would have to be reconsidered at that more fundamental level.
\color{black}

In the regime of interest here, the latter contributions are assumed to represent only a small correction to the explicitly retained current, consistently with the small departures from LCC described by the approximate balance relations derived in \cite{minotti_modanese_math13050892}. These relations take the form of the so-called $\gamma$-model, which will be discussed in more detail in Section \ref{fdt_conductor}.

\color{black}

An alternative scenario can be considered, in which failure of LCC would be of a more fundamental nature. The argument relies on Noether theorem applied to a quantum system that undergoes a collapse of its wavefunction.

\color{black}
For charged fermionic matter described by a globally U(1)-invariant action and evolving on shell according to the corresponding Euler--Lagrange equations, Noether's theorem implies local conservation of the associated matter current.
This statement concerns the global phase symmetry of the matter sector and is distinct from local gauge invariance of the electromagnetic action.
Consequently, if this conserved Noether current is the source of the ABE field equations, the scalar field satisfies the homogeneous equation $\Box S=0$, and Maxwell electrodynamics is recovered for Maxwell-compatible initial and boundary conditions.

\color{black}

The key point is that Noether theorem is valid ``on shell", that is, it requires that the electromagnetic and matter fields evolve according to the equations derived from the action using Hamilton principle. It is argued in \cite{minotti_modanese_epjp2026} that in a quantum collapse process the evolution of the wavefunction does not necessarily follow that derived from Hamilton principle. In particular, at the level of a single-particle Schr\"{o}dinger description, if the collapse of the wavefunction is modeled as an instantaneous transition from the state $\psi_{-}$ to the state $\psi_{+}$, occurring at the time $t=t_{0}$, it is shown in \cite{minotti_modanese_epjp2026} that the charge balance relation results to be
\begin{equation}
    \frac{\partial \rho}{\partial t}+\nabla\cdot {\bf j}=q\left( \left\lvert \psi_{+}\right\lvert^{2}-\left\lvert \psi_{-}\right\lvert^{2} \right)\delta(t-t_{0})
    \label{Noether_collapse}
\end{equation}
where $q$ is the particle charge, and $\delta(...)$ is Dirac's delta.
In systems like biased barriers opposing electron transport, for instance, the generation of an electric current in the circuit sustaining the bias can be interpreted as the collapse of non-localized electron states into states localized at either side of the barrier by tunneling. Remarkably, approximations using relation (\ref{Noether_collapse}) to model this scenario result in a charge balance relation in the form of a $\gamma$-model \cite{minotti_modanese_epjp2026}. 
\color{black}

With the previous considerations, we recall that the Fluctuation Dissipation Theorem (FDT) \cite{callen1951} can be used to determine the equilibrium state of electromagnetic radiation interacting with matter in equilibrium with a thermal bath.

In particular, if one considers Maxwell equations with material sources that are charges and currents without any additional restriction than satisfying LCC and being in thermal equilibrium, the standard black-body spectrum is obtained, including the contribution of the zero-point energy. If the material sources are further constrained, for instance by imposing constitutive relations in a medium, like a conductor, expressions of the Johnson-Nyquist thermal noise can be determined. \color{black}These known results are recovered in the following Sections. \color{black} 

In this way, it is of interest to explore which would be the results of applying the FDT to a theory like ABE, which can describe the electromagnetic field generated by sources that do not satisfy strict LCC.  \color{black} Conditions for the applicability of the FDT to ABE, which are essentially the same as those required for its use in Maxwell electrodynamics, are considered in the next section. \color{black}

In the application of the FDT to ABE the non enforcing of LCC results in the impossibility of fixing the gauge. This has the effect of increasing the degree of freedom of the variables describing the system, with the consequence that, at variance with Maxwell case, the contribution of electric and magnetic fields to the energy spectrum is not the same, although the spectrum of the total energy coincides with Planck's expression, plus the contribution of the zero-point energy.

Similarly, additional effects are obtained for a conducting medium with no strict local conservation of charge, described by the so called ``$\gamma$-model" \cite{minotti_modanese_epjc2023,minotti_modanese_math13050892}, resulting in the addition of a colored contribution to the classical white-noise Johnson-Nyquist result.   

In the next section a brief summary of ABE is presented, remarking the particular features of the theory required for the application of the FDT. In Section \ref{fdt_ab} the FDT is recalled, and applied to a system in which LCC does not hold. In Section \ref{fdt_maxwell} we consider a system with strict LCC, while in Section \ref{fdt_conductor} we study electromagnetic fluctuations in a material that admits both kinds of sources, as described by a simple constitutive model. In Section \ref{tests} we discuss the possibility of an experimental test of the results in the previous section. The last section contains some conclusions and further considerations.

\section{Summary of Aharonov–Bohm electrodynamics}
\label{AB_summary}

Aharonov–Bohm electrodynamics (ABE) extends classical electrodynamics by allowing, in principle, for local non-conservation of charge. In this framework the quantity
\begin{equation}
    I=\frac{\partial \rho}{\partial t}+\nabla\cdot {\bf j}
    \label{ABI}
\end{equation}
needs not vanish identically and measures the degree of local non-conservation of the material sources. A convenient scalar quantity associated with the potentials is
\begin{equation}
    S=\frac{1}{c^2}\frac{\partial \phi}{\partial t}+\nabla\cdot {\bf A},
    \label{ABscalar}
\end{equation}
which in Maxwell electrodynamics vanishes when the Lorentz gauge is imposed on strictly conserved sources, but is in general non-zero in ABE.

\color{black}
ABE is derived from an action for the electromagnetic field that is identical to that corresponding to Maxwell electrodynamics, but that includes a ``gauge-fixing term". In four-vector notation, with $A^{i}=(\phi/c,\bf A)$, the gauge-fixing term that is added to Maxwell's Lagrangian density is of the form
\begin{equation}
    L_{GF}=-\frac{1}{2\mu_{0}}(\partial_{i}A^{i})^{2},
    \label{gauge_fixing}
\end{equation}
which in the context of QED corresponds to what is sometimes called the Feynman gauge. 
A relevant point for the application of the FDT to ABE is that, as Maxwell electrodynamics, ABE has a Hamiltonian structure that can be derived from its Lagrangian density. In fact, one of the reasons for the introduction of the particular gauge-fixing term (\ref{gauge_fixing}) in QED is to ensure that the canonical momentum conjugate to the scalar potential $\phi$ is not identically zero.

In the context of classical electrodynamics the interaction term in the Lagrangian density is also the same as Maxwell's, $-J_{i}A^{i}$, with the four-current $J_{i}=(\rho c,-\bf j)$. 
In this way, variation of $A^{i}$ in the total action finally results in the same homogeneous equations as Maxwell's
\begin{equation}
    \nabla\times{\bf E }=-\frac{\partial {\bf B}}{\partial t}, \qquad \nabla \cdot{\bf B} =0,
    \label{AB_fields_ho}
\end{equation}
where electromagnetic fields and potentials are related in the usual way,
\begin{equation}
    {\bf E}=-\frac{\partial\bf A}{\partial t}-\nabla{ \phi},\qquad {\bf B}=\nabla \times {\bf A}.
    \label{fields_potentials}
\end{equation}
The non-homogeneous equations are
\begin{equation}
    \frac{1}{c^2}\frac{\partial^2 \phi}{\partial t^2}-\nabla^2\phi=\frac{\rho}{\varepsilon_0},
\qquad
\frac{1}{c^2}\frac{\partial^2 {\bf A}}{\partial t^2}-\nabla^2{\bf A}=\mu_0 {\bf j},
\label{ABpotentials}
\end{equation}
with retarded solutions formally identical to those obtained in Maxwell electrodynamics in the Lorentz gauge. The conceptual difference is that, in ABE, these equations do not rely on strict local conservation of charge as an a priori condition, and therefore the scalar and vector sectors are not constrained in the same way as in Maxwell theory.

From relations (\ref{fields_potentials}) and (\ref{ABpotentials}) we can write the non-homogeneous equations for the fields as
\begin{equation}
    \nabla \cdot \mathbf{E} =\frac{\rho }{\varepsilon _{0}}-\frac{\partial S}{
\partial t},
\qquad
\nabla \times \mathbf{B} =\mu _{0}\mathbf{j}+\frac{1}{c^{2}}\frac{\partial 
\mathbf{E}}{\partial t}+\nabla S,
\label{AB_fields_nh}
\end{equation}
which show how the scalar $S$ acts as an effective source for the electric and magnetic fields.

In addition, from the expression (\ref{ABscalar}) of the ABE scalar $S$, and relations (\ref{ABpotentials}), one easily obtains 
\begin{equation}
    \frac{1}{c^2}\frac{\partial^2 S}{\partial t^2}-\nabla^2 S=\mu_0I,
    \label{equation_S}
\end{equation}
with $I$ defined by (\ref{ABI}). In this way, the charge and current densities in the regions where LCC fails act as material sources of $S$.

\color{black}
The scalar field $S$ therefore represents a genuine additional dynamical sector of ABE. It should not, however, be interpreted as an additional polarization necessarily accompanying ordinary electromagnetic radiation. For locally conserved sources, $I=0$, and with Maxwell-compatible initial and boundary conditions (in particular, in the absence of an independently incoming homogeneous scalar wave), $S=0$, so that the usual two transverse radiative polarizations are recovered.
\color{black}

\color{black}
The fact that the scalar sector cannot in general be eliminated as a gauge artifact when $I\neq 0$
is directly related to the gauge properties of the source--field interaction.
Indeed the interaction term $S_{\rm int}=-\int d^4xJ^i A_i$ changes under a variation $A_i\rightarrow A_i+\partial_i\chi$ (apart from a border term), as
$\delta S_{\rm int}=\int d^4x\chi\partial_i J^i$. Therefore, if $\partial_i J^i\neq 0$,
an arbitrary local gauge transformation does not leave the coupled system invariant and in general is not possible to ``gauge away" $S$. 
\color{black}

Furthermore, by writing the ABE action in a general four-dimensional metric, the ABE energy-momentum tensor can be determined by variation of the metric \cite{Landau_fields}, and the expression of the conservation of energy obtained from the four-divergence of the resulting energy-momentum tensor \cite{minotti2021quantum}.
The relevant result is that the local balance equation for the electromagnetic energy density $u$ can be written as
\begin{equation}
    \frac{\partial u}{\partial t}+\nabla\cdot {\bf S}_u+{\bf j}\cdot{\bf E}-I\phi=0.
    \label{energy_conservation}
\end{equation}
The energy density $u$ contains, besides the standard electric and magnetic contributions, additional terms involving the scalar quantity $S$. In particular,
\begin{equation}
    u=\frac{1}{\mu_0}\left(
\frac{|{\bf E}|^2}{2c^2}
+\frac{|{\bf B}|^2}{2}
+\frac{\phi}{c^2}\frac{\partial S}{\partial t}
-{\bf A}\cdot\nabla S
-\frac{S^2}{2}
\right).
\label{AB_u}
\end{equation}
The expression of the generalized Poynting vector is
\begin{equation}
    {\bf S}_u=\frac{1}{\mu_{0}}\left({\bf E} \times {\bf B} -\phi\nabla S+{\bf A}\frac{\partial S}{\partial t}  \right).
    \label{AB_Su}
\end{equation}

A property of the energy density (\ref{AB_u}) is that it is not positive definite, which presents a potential danger for the stability of ABE. There is to our knowledge no general proof of which are the allowed physical systems described by ABE. Examples of sources that violated LCC were analyzed in \cite{minotti2021quantum}, with the conclusion that some were not physically possible, based on the impossibility of setting them up, even using non-electromagnetic actions. In particular, a time-varying dipole of pure charges, without currents, whose radiation field has negative energy, was determined to be physically impossible.

\color{black}
An argument in favor of the stability of ABE is that relation (\ref{energy_conservation}) can be recast as \cite{minotti2021quantum}
\begin{equation}
\frac{\partial \widetilde{u}}{\partial t}+\nabla \cdot \widetilde{\mathbf{S}}
_{u}+\mathbf{j}\cdot \mathbf{E}-c^{2}\rho S=0,
\label{energy_mod}
\end{equation}
where
\begin{eqnarray*}
\widetilde{u} &=&\frac{1}{\mu _{0}}\left( \frac{\left\vert \mathbf{E}
\right\vert ^{2}}{2c^{2}}+\frac{\left\vert \mathbf{B}\right\vert ^{2}}{2}+
\frac{S^{2}}{2}\right) , \\
\widetilde{\mathbf{S}}_{u} &=&\frac{1}{\mu _{0}}\left( \mathbf{E}\times 
\mathbf{B}+S\mathbf{E}\right). 
\end{eqnarray*}
Although $\widetilde{u}$ is positive definite one must analyze how the modified Poynting vector $\widetilde{\mathbf{S}}_{u}$ acts, in order to ascertain how the electromagnetic fields localized in a given region evolve. For this we analyze the radiative field away from the material sources in that region and consider the plane wave components of angular frequency $\omega $ of the fields $S$, $\mathbf{E}$ and $\mathbf{B}$, written generically as the real parts of the expressions
\begin{eqnarray*}
S_{0}\exp \left[ i\left( \mathbf{k}\cdot \mathbf{x}-\omega t\right) 
\right] , \\
\mathbf{E}_{0}\exp \left[ i\left( \mathbf{k}\cdot \mathbf{x}
-\omega t\right) \right] , \\
\mathbf{B}_{0}\exp \left[ i\left( \mathbf{k}\cdot \mathbf{x}
-\omega t\right) \right] ,
\end{eqnarray*}
with constant complex amplitudes $S_{0}$, $\mathbf{E}_{0}$ and $\mathbf{B}_{0}$, and
where the wavevector $\mathbf{k}$ satisfies $\left\vert \mathbf{k}
\right\vert\equiv k =\omega /c$.

From equations (\ref{AB_fields_ho}) and (\ref{AB_fields_nh}) we thus have that 
\begin{equation*}
S_{0}=\frac{\mathbf{k}\cdot \mathbf{E}_{0}}{\omega },\qquad \mathbf{B}_{0}=\frac{\mathbf{k}\times \mathbf{E}_{0}}{\omega }.
\end{equation*}
Note that, at variance with Maxwell electrodynamics, $\mathbf{E}_{0}$ can have a component parallel to $\mathbf{k}$.

The time-averaged value of the component of $\widetilde{\mathbf{S}}_{u}$
with frequency $\omega $ can thus be expressed as ($\langle ...\rangle$ represents the time average)
\begin{equation}
\langle \widetilde{\mathbf{S}}_{u}\rangle =\frac{1}{2\mu
_{0}\omega } \left\vert 
\mathbf{E}_{0}\right\vert ^{2} \mathbf{k} .
\label{Su_mod}
\end{equation}

Considering the flow of $\langle \widetilde{\mathbf{S}}_{u}\rangle$ across the boundary of the region containing the fields, we see from (\ref{energy_mod}) and (\ref{Su_mod}) that for outgoing radiation the value of the positive-definite $\widetilde{u}$ in the interior decreases. The material sources $\rho$ and $\mathbf{j}$, on the other hand, act locally, according to (\ref{energy_mod}), either dissipating or enhancing the fields energy, according to the particular dynamics of those sources, which, as detailed in \cite{minotti2021quantum}, is derived from the local conservation of the total, fields plus matter, energy-momentum. 
These considerations suggest that no run-away behavior should occur for the class of radiative configurations considered here, although they do not constitute a general proof of stability for arbitrary ABE sources.
\color{black}

Furthermore, equation (\ref{equation_S}) suggests that, for sufficiently small violations of LCC as quantified by $I$, the scalar sector excited by the material sources may also remain perturbatively small in physically relevant configurations. This is consistent with the above mentioned $\gamma$-model, in which the relative magnitude of $I$ is quantified in terms of the non-dimensional parameter $\gamma$, which in the many-body systems studied that could present a failure of LCC has an estimated magnitude much smaller than one.

In order to explore further this issue we study the ABE radiation field, considering, as we did above, a plane-wave component of angular frequency $\omega $ of the solution of (\ref{ABpotentials}) outside the material sources, written as the real parts of the expressions
\begin{eqnarray*}
\phi  = \phi _{0}\exp[i (\mathbf{k}\cdot \mathbf{x}-\omega t)]
,\qquad
\mathbf{A} =\mathbf{A}_{0}\exp[i (\mathbf{k}\cdot \mathbf{x}
-\omega t)] ,
\end{eqnarray*}
where $\phi _{0}$ and $\mathbf{A}_{0}$ are constant complex amplitudes.

By further decomposing $\mathbf{A}_{0}$ in a component parallel to $\mathbf{k}$, $A_{0\Vert }$, and a perpendicular vector component, $\mathbf{A}_{0\bot }$, the time-averaged terms of the corresponding
energy density expression (\ref{AB_u}) are readily evaluated to be (Re(...) represents the real part)
\begin{eqnarray*}
\frac{\left\langle \left\vert \mathbf{E}\right\vert ^{2}\right\rangle }{2\mu
_{0}c^{2}} &=&\frac{k^{2}}{4\mu _{0}}\left( \left\vert A_{0\Vert }-\frac{
\phi _{0}}{c}\right\vert ^{2}+\left\vert \mathbf{A}_{0\bot }\right\vert
^{2}\right) , \\
\frac{\left\langle \left\vert \mathbf{B}\right\vert ^{2}\right\rangle }{2\mu
_{0}} &=&\frac{k^{2}}{4\mu _{0}}\left\vert \mathbf{A}_{0\bot }\right\vert
^{2}, \\
\frac{1}{\mu _{0}c^{2}}\left\langle \phi \frac{\partial S}{\partial t}
\right\rangle  &=&\frac{1}{2\mu _{0}}\left[ \text{Re}\left( k^{2}\frac{\phi
_{0}^{\ast }}{c}A_{0\Vert }\right) -\frac{k^{2}}{c^{2}}\left\vert \phi
_{0}\right\vert ^{2}\right] , \\
\frac{1}{\mu _{0}}\left\langle \mathbf{A}\cdot \nabla S\right\rangle  &=&
\frac{1}{2\mu _{0}}\left[ \text{Re}\left( k^{2}\frac{\phi _{0}^{\ast }}{c}
A_{0\Vert }\right) -k^{2}\left\vert A_{0\Vert }\right\vert ^{2}\right] , \\
\frac{\left\langle S^{2}\right\rangle }{2\mu _{0}} &=&\frac{k^{2}}{4\mu _{0}}
\left\vert A_{0\Vert }-\frac{\phi _{0}}{c}\right\vert ^{2}.
\end{eqnarray*}

The time-averaged energy density of the $\omega $ component thus results in
\begin{equation}
\left\langle u\right\rangle =\frac{k^{2}}{2\mu _{0}}\left( \left\vert 
\mathbf{A}_{0\bot }\right\vert ^{2}+\left\vert A_{0\Vert }\right\vert ^{2}-
\frac{\left\vert\phi _{0}\right\vert ^{2}}{c^{2}}\right) .
\label{u_radiation}
\end{equation}

A first thing to note is that the magnitude of the electric contribution is
equal to that of the magnetic one, plus that of the scalar contribution $
\left\langle S^{2}\right\rangle /\left( 2\mu _{0}\right) $, a result that
will be recovered in the next section in the context of the FDT.

\color{black}
Considering that the complex amplitude of the $\omega$ component of $S$ is
\begin{equation}
S_0=ik\left(A_{0\Vert}-\frac{\phi_0}{c}\right),
\end{equation}
the magnitude of the scalar field is quantified by
\begin{equation}
|S_0|^2
=
k^2\left|A_{0\Vert}-\frac{\phi_0}{c}\right|^2.
\end{equation}
This condition should be distinguished from that determining the sign of the
time-averaged energy density. From Eq.\ \eqref{u_radiation}, positivity of
$\langle u\rangle$ requires
\begin{equation}
|\mathbf{A}_{0\bot}|^2
+
|A_{0\Vert}|^2
-
\frac{|\phi_0|^2}{c^2}
>0.
\end{equation}
Thus, smallness of $S$ and positivity of the energy density are in general
distinct conditions, since the former also depends on the relative phase
between $A_{0\Vert}$ and $\phi_0$. In particular, Maxwell electrodynamics is
recovered when $S=0$, corresponding to the equality of the complex amplitudes
\begin{equation}
A_{0\Vert}=\frac{\phi_0}{c}.
\end{equation}

For the limiting case considered in the next section, in which no local
charge-balance constraint is imposed, application of the FDT gives
\begin{equation}
|A_{0\Vert}|^2=\frac{|\phi_0|^2}{c^2}.
\end{equation}
These two terms therefore cancel in Eq.~\eqref{u_radiation}, leaving
\begin{equation}
\langle u\rangle
=
\frac{k^2}{2\mu_0}|\mathbf{A}_{0\bot}|^2,
\end{equation}
which is non-negative and coincides with the Maxwell result. Notice that this
cancellation does not imply $S=0$: in this case the FDT correlations instead
give
\begin{equation}
\left|A_{0\Vert}-\frac{\phi_0}{c}\right|
=
|\mathbf{A}_{0\bot}|,
\end{equation}
so that the scalar sector remains non-vanishing in general.

Although these considerations do not constitute a general proof that the full ABE Hamiltonian is bounded from below for arbitrary sources, they show that the particular equilibrium sector considered below has a non-negative mean energy. Our application of equilibrium statistical mechanics is therefore restricted to such physically stable configurations. The general stability problem of ABE with arbitrary non-conserved sources remains open.

\color{black}

\color{black}
Finally, after integration of relation (\ref{energy_conservation}) over a volume including all sources, assumed to be localized, one obtains (as detailed in the next section)
\begin{equation}
    \frac{dU}{dt}+\oint {\bf S}_u\cdot d{\bf S}=
\int\left(\phi\frac{\partial \rho}{\partial t}
+{\bf j}\cdot\frac{\partial {\bf A}}{\partial t}\right)dV.
\end{equation}
This expression is especially convenient for the application of the FDT, because it directly identifies the relevant dynamical variables and their conjugate generalized forces. \color{black} It also shows that the action of the generalized forces is expressed through a linear functional, a condition for the applicability of the FDT, and which in addition has the same expression as in Maxwell theory.   \color{black}

As a consequence, even when the total equilibrium spectral energy density retains the standard Planck form, its partition among electric, magnetic, and scalar contributions may differ from the Maxwell case. This is precisely the aspect that will be relevant in the fluctuation analysis below.

In the following, this structure will be used to apply the fluctuation-dissipation theorem first to a system of non-conserved sources, and then to a conducting medium described by the $\gamma$-model.


\color{black}
The application of the FDT below concerns a stationary ensemble in thermal equilibrium, rather than the dynamics of individual microscopic events that may generate a non-vanishing $I$. We assume that the macroscopic linear response of the equilibrium system satisfies microscopic reversibility and that the relevant equilibrium sector is stable. Under these assumptions, the generalized susceptibilities introduced below retain their standard linear-response interpretation.
\color{black}

\section{ABE fluctuations with non-conserved sources}
\label{fdt_ab}

In this section we consider a system in which charge and current densities, $\rho$ and $\mathbf{j}$, respectively, which are the material sources of the electromagnetic field, are not constrained to satisfy strict local conservation, that is, $\partial\rho/\partial t+\nabla\cdot\mathbf{j}$ is generally not zero, and has no a priori imposed values. In principle, \color{black} as discussed in the Introduction\color{black},  this situation could only arise in many-body systems with strong interactions, and only in regions where collective quantum effects allows it. In this way, one expects that in an actual situation the contribution to the electromagnetic fluctuations of these non-conserved sources constitute a (possibly small) fraction of the fluctuations due to conserved sources, which can be described by Maxwell electrodynamics. In order to quantify the electromagnetic fluctuations in a material system that includes both kinds of sources, a constitutive relation that relates them is needed. An attempt on this line is done in Section \ref{fdt_conductor}, where a simple constitutive model is applied to determine electromagnetic fluctuations in a conductor. In the present section and the next one we consider only the cases in which either non-conservation or strict conservation applies to the whole system.

To begin, we briefly recall the FDT in the form to be applied below, for which we essentially follow \cite{Landau_Stat_Phys}. 

Consider a system described by the discrete variables $x_{a}\left( t\right) $. If acted by an external perturbation, with forces $f_{a}\left( t\right) $, the system energy $U$ is modified at a rate
\begin{equation}
\frac{dU}{dt}=-\sum\limits_{a}x_{a}\left( t\right) \frac{df_{a}}{dt}.  \label{dEdt}
\end{equation}

Furthermore, in the frequency domain the relation between variables $x_{a}\left(\omega\right)$ and the external forces is expressed, using the generalized susceptibility tensor $\alpha _{ab}\left( \omega \right)$, as
\begin{equation}
x_{a}\left( \omega \right) =\sum\limits_{b}\alpha _{ab}\left( \omega \right) f_{b}\left(
\omega \right) .  \label{xaf}
\end{equation}

The FDT then says that in equilibrium the spectral distribution of fluctuations of the $x_{a}$'s is given as
\begin{equation}
\left( x_{a}x_{b}\right) _{\omega }=\frac{i\hbar }{2}\left[ \alpha
_{ba}^{\ast }\left( \omega \right) -\alpha _{ab}\left( \omega \right) \right]
\coth \left( \frac{\hbar \omega }{2T}\right) ,  \label{xaxb}
\end{equation}
with the definitions (the brackets represent average over the fluctuations)
\begin{eqnarray}
\left( x_{a}x_{b}\right) _{\omega } &\equiv &\int_{-\infty }^{+\infty
}\left\langle x_{a}\left( t\right) x_{b}\left( 0\right) \right\rangle
e^{i\omega t}dt,  \label{xaxbw} \\
\left\langle x_{a}\left( \omega \right) x_{b}\left( \omega ^{\prime }\right)
\right\rangle &=&2\pi \left( x_{a}x_{b}\right) _{\omega }\delta \left(
\omega +\omega ^{\prime }\right) .  \label{xawxbwp}
\end{eqnarray}

On the other hand, since
\begin{equation}
f_{a}\left( \omega \right) =\sum\limits_{b}\alpha _{ab}^{-1}\left( \omega \right)
x_{b}\left( \omega \right) ,  \label{alfainv}
\end{equation}
one has that
\begin{equation*}
\left( f_{a}f_{b}\right) _{\omega }=\sum\limits_{k,m}\alpha _{ak}^{-1}\left( \omega \right)
\alpha _{bm}^{-1}\left( \omega \right) \left( x_{k}x_{m}\right) _{\omega },
\end{equation*}
which results in
\begin{equation}
\left( f_{a}f_{b}\right) _{\omega }=\frac{i\hbar }{2}\left[ \alpha
_{ab}^{-1}\left( \omega \right) -\alpha _{ba}^{-1\ast }\left( \omega \right) 
\right] \coth \left( \frac{\hbar \omega }{2T}\right) .  \label{fafb}
\end{equation}

In order to apply the previous relations to ABE, we integrate relation (\ref{energy_conservation}) over a space volume that includes all sources, which are assumed to be localized, to obtain 
\begin{equation*}
\frac{dU}{dt}+\int \mathbf{S}_{u}\cdot d\mathbf{S}=-\int \left( \mathbf{j}
\cdot \mathbf{E}-I\phi \right) dV.
\end{equation*}

We first note that
\begin{eqnarray*}
\mathbf{j}\cdot \mathbf{E}-I\phi &=&\mathbf{j}\cdot \left( -\nabla \phi -
\frac{\partial \mathbf{A}}{\partial t}\right) -\left( \frac{\partial \rho }{
\partial t}+\nabla \cdot \mathbf{j}\right) \phi \\
&=&\mathbf{j}\cdot \left( -\nabla \phi -\frac{\partial \mathbf{A}}{\partial t
}\right) -\frac{\partial \rho }{\partial t}\phi -\nabla \cdot \left( \mathbf{
j}\phi \right) +\mathbf{j}\cdot \nabla \phi \\
&=&-\phi \frac{\partial \rho }{\partial t}-\mathbf{j}\cdot \frac{\partial 
\mathbf{A}}{\partial t}-\nabla \cdot \left( \mathbf{j}\phi \right) .
\end{eqnarray*}
Since the volume of integration includes all sources, one has at the surface 
$\mathbf{j}\cdot d\mathbf{S}=0$, which allows us to finally write
\begin{equation}
\frac{dU}{dt}+\int \mathbf{S}_{u}\cdot d\mathbf{S}=\int \left( \phi \frac{
\partial \rho }{\partial t}+\mathbf{j}\cdot \frac{\partial \mathbf{A}}{
\partial t}\right) dV.  \label{dEdtfin}
\end{equation}

For the application of the FDT, one can consider a discrete set of small
volumes $\triangle V$, at spatial positions $\mathbf{r}_{a}$, as done in \cite{Landau_ED_Cont_Media}, and, by comparison of relation (\ref{dEdtfin}) with (\ref{dEdt}), identify as the discrete variables 
\begin{eqnarray*}
x_{a}\left( t\right) &\rightarrow &\phi \left( \mathbf{r}_{a},t\right) ,
\text{ }j_{\beta }\left( \mathbf{r}_{a},t\right) \triangle V, \\
f_{a}\left( t\right) &\rightarrow &-\rho \left( \mathbf{r}_{a},t\right)
\triangle V,\text{ }-A_{\beta }\left( \mathbf{r}_{a},t\right) ,
\end{eqnarray*}
where Greek indices correspond to Cartesian vector components.

Further, as mentioned above, the equations for the potentials in ABE coincide with those in Maxwell electrodynamics in the Lorentz gauge \cite{minotti2021quantum} 
\begin{subequations}
\label{potwaves}
\begin{eqnarray}
\frac{1}{c^{2}}\frac{\partial ^{2}\phi }{\partial t^{2}}-\nabla ^{2}\phi &=&
\frac{\rho }{\varepsilon _{0}},  \label{phiwave} \\
\frac{1}{c^{2}}\frac{\partial ^{2}A_{\beta }}{\partial t^{2}}-\nabla
^{2}A_{\beta } &=&\mu _{0}j_{\beta },  \label{Awave}
\end{eqnarray}
with general solution, for localized sources in unlimited space,
\end{subequations}
\begin{eqnarray*}
\phi \left( \mathbf{r},t\right) &=&\frac{1}{4\pi \varepsilon _{0}}\int \frac{
\rho \left( \mathbf{r}^{\prime },t^{\prime }\right) }{\left\vert \mathbf{r}-
\mathbf{r}^{\prime }\right\vert }dV^{\prime }, \\
A_{\beta }\left( \mathbf{r},t\right) &=&\frac{\mu _{0}}{4\pi }\int \frac{
j_{\beta }\left( \mathbf{r}^{\prime },t^{\prime }\right) }{\left\vert 
\mathbf{r}-\mathbf{r}^{\prime }\right\vert }dV^{\prime },
\end{eqnarray*}
where $t^{\prime }=t-\left\vert \mathbf{r}-\mathbf{r}^{\prime }\right\vert /c$.

We can thus write in the frequency domain $\phi \left( \mathbf{r}^{\prime
},t^{\prime }\right) \rightarrow \phi \left( \mathbf{r}^{\prime },\omega
\right) \exp \left( -i\omega t^{\prime }\right) $
\begin{equation*}
\phi \left( \mathbf{r},\omega \right) =\frac{1}{4\pi \varepsilon _{0}}\int 
\frac{\rho \left( \mathbf{r}^{\prime },\omega \right) }{\left\vert \mathbf{r}
-\mathbf{r}^{\prime }\right\vert }\exp \left( i\omega \left\vert \mathbf{r}-
\mathbf{r}^{\prime }\right\vert /c\right) dV^{\prime },
\end{equation*}
which in discrete form is
\begin{equation}
\phi \left( \mathbf{r}_{a},\omega \right) =\frac{1}{4\pi \varepsilon _{0}}\sum\limits_{b}
\frac{\exp \left( i\omega \left\vert \mathbf{r}_{a}-\mathbf{r}
_{b}\right\vert /c\right) }{\left\vert \mathbf{r}_{a}-\mathbf{r}
_{b}\right\vert }\rho \left( \mathbf{r}_{b},\omega \right) \triangle V.
\label{phiaw}
\end{equation}
Analogously
\begin{equation}
A_{\beta }\left( \mathbf{r}_{a},\omega \right) =\frac{\mu _{0}}{4\pi }\sum\limits_{b}\frac{
\exp \left( i\omega \left\vert \mathbf{r}_{a}-\mathbf{r}_{b}\right\vert
\right) /c}{\left\vert \mathbf{r}_{a}-\mathbf{r}_{b}\right\vert }j_{\beta
}\left( \mathbf{r}_{b},\omega \right) \triangle V.  \label{Aaw}
\end{equation}

Expressions (\ref{phiaw}) and (\ref{Aaw}) allow to easily determine the tensor $\alpha _{ab}$. First note that $\phi \left( \mathbf{r}_{a},\omega
\right) $ is only related to the elements $\rho \left( \mathbf{r}_{b},\omega
\right) \triangle V$, and that each Cartesian component $A_{\beta }\left( 
\mathbf{r}_{a},\omega \right) $ to only the corresponding component of the
elements $j_{\beta }\left( \mathbf{r}_{b},\omega \right) \triangle V$. In
this way, the tensor $\alpha _{ab}$ breaks down into four independent parts.
For the part relating $x_{a}\left( \omega \right) \equiv \phi \left( \mathbf{
r}_{a},\omega \right) $ to $f_{a}\left( \omega \right) \equiv -\rho \left( 
\mathbf{r}_{a},\omega \right) \triangle V$, we have from (\ref{phiaw})
\begin{equation*}
\alpha _{ab}\left( \omega \right) =-\frac{1}{4\pi \varepsilon _{0}}\frac{
\exp \left( i\omega \left\vert \mathbf{r}_{a}-\mathbf{r}_{b}\right\vert
/c\right) }{\left\vert \mathbf{r}_{a}-\mathbf{r}_{b}\right\vert }.
\end{equation*}

From (\ref{Aaw}), for the tensor part corresponding to each Cartesian component $\beta $ relating $x_{a}\left( \omega \right) \equiv $ $j_{\beta }\left( 
\mathbf{r}_{a},\omega \right) \triangle V$ and $f_{a}\left( \omega \right)
\equiv -A_{\beta }\left( \mathbf{r}_{a},\omega \right) $, using (\ref{alfainv}),
\begin{equation*}
\alpha _{ab}^{-1}\left( \omega \right) =-\frac{\mu _{0}}{4\pi }\frac{\exp
\left( i\omega \left\vert \mathbf{r}_{a}-\mathbf{r}_{b}\right\vert /c\right) 
}{\left\vert \mathbf{r}_{a}-\mathbf{r}_{b}\right\vert }.
\end{equation*}

We thus have, from (\ref{xaxb}),
\begin{equation}
\left( \phi ^{\mathbf{r}_{a}}\phi ^{\mathbf{r}_{b}}\right) _{\omega }=-\frac{
\hbar }{4\pi \varepsilon _{0}}\frac{\sin \left( \omega \left\vert \mathbf{r}
_{a}-\mathbf{r}_{b}\right\vert /c\right) }{\left\vert \mathbf{r}_{a}-\mathbf{
r}_{b}\right\vert }\coth \left( \frac{\hbar \omega }{2T}\right) ,
\label{phiaphibw}
\end{equation}
while, from (\ref{fafb}),
\begin{equation}
\left( A_{\beta }^{\mathbf{r}_{a}}A_{\gamma }^{\mathbf{r}_{b}}\right)
_{\omega }=\frac{\hbar \mu _{0}}{4\pi }\frac{\sin \left( \omega \left\vert 
\mathbf{r}_{a}-\mathbf{r}_{b}\right\vert /c\right) }{\left\vert \mathbf{r}
_{a}-\mathbf{r}_{b}\right\vert }\coth \left( \frac{\hbar \omega }{2T}\right)
\delta _{\beta \gamma }.  \label{AbAg}
\end{equation}

Besides, due to the non existence of crossed terms, we have $\left( \phi ^{
\mathbf{r}_{a}}A_{\beta }^{\mathbf{r}_{b}}\right) _{\omega }=0$.

We can thus write 
\begin{subequations}
\label{correlfull}
\begin{eqnarray}
\left\langle \phi \left( \mathbf{r}_{a},\omega \right) A_{\beta }\left( 
\mathbf{r}_{b},\omega ^{\prime }\right) \right\rangle &=&0, \\
\left\langle \phi \left( \mathbf{r}_{a},\omega \right) \phi \left( \mathbf{r}
_{b},\omega ^{\prime }\right) \right\rangle &=&F_{ab}\left( \omega \right)
\coth \left( \frac{\hbar \omega }{2T}\right) \delta \left( \omega +\omega
^{\prime }\right) , \\
\left\langle A_{\beta }\left( \mathbf{r}_{a},\omega \right) A_{\gamma
}\left( \mathbf{r}_{b},\omega ^{\prime }\right) \right\rangle &=&\frac{
F_{ab}\left( \omega \right) }{c^{2}}\coth \left( \frac{\hbar \omega }{2T}
\right) \delta \left( \omega +\omega ^{\prime }\right) \delta _{\beta \gamma},
\end{eqnarray}
with 
\end{subequations}
\begin{equation*}
F_{ab}\left( \omega \right) =\frac{\hbar }{2\varepsilon _{0}}\frac{\sin
\left( \omega \left\vert \mathbf{r}_{a}-\mathbf{r}_{b}\right\vert /c\right) 
}{\left\vert \mathbf{r}_{a}-\mathbf{r}_{b}\right\vert },
\end{equation*}
and where the relation $\mu _{0}\varepsilon _{0}c^{2}=1$\ was used.

From the previous expressions we can now determine the spectrum of mean energy density of the fluctuations. We recall the expression of the energy density $u$ in ABE,
\begin{equation}
u=\frac{1}{\mu _{0}}\left( \frac{\left\vert \mathbf{E}\right\vert ^{2}}{
2c^{2}}+\frac{\left\vert \mathbf{B}\right\vert ^{2}}{2}+\frac{\phi }{c^{2}}
\frac{\partial S}{\partial t}-\mathbf{A}\cdot \nabla S-\frac{S^{2}}{2}
\right) ,
\end{equation}
where
\begin{equation}
S=\frac{1}{c^{2}}\frac{\partial \phi }{\partial t}+\nabla \cdot \mathbf{A},
\end{equation}

Using the correlations (\ref{correlfull}) we have that
\begin{eqnarray*}
\left\langle \frac{\phi }{c^{2}}\frac{\partial S}{\partial t}\right\rangle 
&=&\frac{1}{c^{4}}\left\langle \phi \frac{\partial ^{2}\phi }{\partial t^{2}}
\right\rangle  \\
&=&-\frac{1}{\left( 2\pi \right) ^{2}c^{4}}\int_{-\infty }^{\infty
}\int_{-\infty }^{\infty }\omega ^{\prime 2}\left\langle \phi ^{\mathbf{r}
_{a}}\left( \omega \right) \phi ^{\mathbf{r}_{a}}\left( \omega ^{\prime
}\right) \right\rangle e^{-i\left( \omega +\omega ^{\prime }\right)
t}d\omega d\omega ^{\prime } \\
&=&-\frac{1}{\left( 2\pi \right) ^{2}}\int_{-\infty }^{\infty }\frac{\hbar
\omega ^{3}}{2\varepsilon _{0}c^{5}}\coth \left( \frac{\hbar \omega }{2T}
\right) d\omega ,
\end{eqnarray*}
and
\begin{eqnarray*}
\left\langle \mathbf{A}\cdot \nabla S\right\rangle  &=&\left\langle \mathbf{A
}\cdot \nabla \left( \nabla \cdot \mathbf{A}\right) \right\rangle  \\
&=&\lim_{\mathbf{r}_{b}\rightarrow \mathbf{r}_{a}}\frac{1}{\left( 2\pi
\right) ^{2}}\int_{-\infty }^{\infty }\int_{-\infty }^{\infty }\frac{
\partial ^{2}}{\partial r_{\beta }^{b}\partial r_{\gamma }^{b}}\left\langle
A_{\beta }^{\mathbf{r}_{a}}\left( \omega \right) A_{\gamma }^{\mathbf{r}
_{b}}\left( \omega ^{\prime }\right) \right\rangle e^{-i\left( \omega
+\omega ^{\prime }\right) t}d\omega d\omega ^{\prime } \\
&=&-\frac{1}{\left( 2\pi \right) ^{2}}\int_{-\infty }^{\infty }\frac{\hbar
\mu _{0}\omega ^{3}}{2c^{3}}\coth \left( \frac{\hbar \omega }{2T}\right)
d\omega ,
\end{eqnarray*}
where we used the property of correlations (\ref{correlfull}) ($r_{\beta }^{a}$ represents the $\beta$ component of the vector $\mathbf{r}_{a}$)  
\begin{equation*}
\frac{\partial
\left\langle ...\right\rangle }{\partial r_{\beta }^{b}}=-\frac{\partial
\left\langle ...\right\rangle} {\partial r_{\beta }^{a}},
\end{equation*}
together with the identity
\begin{equation*}
\left[ \frac{\omega ^{2}}{c^{2}}+\nabla _{a}^{2}\right] \frac{\sin \left(
\omega \left\vert \mathbf{r}_{a}-\mathbf{r}_{b}\right\vert /c\right) }{
\left\vert \mathbf{r}_{a}-\mathbf{r}_{b}\right\vert }=0.
\end{equation*}

We thus see that for the fluctuations considered 
\begin{equation*}
\left\langle \frac{\phi }{c^{2}}\frac{\partial S}{\partial t}-\mathbf{A}
\cdot \nabla S\right\rangle =0.
\end{equation*}

Similarly, we have 
\begin{eqnarray*}
\left\langle S^{2}\right\rangle  &=&\frac{1}{c^{4}}\left\langle \left( \frac{
\partial \phi }{\partial t}\right) ^{2}\right\rangle +\left\langle \left(
\nabla \cdot \mathbf{A}\right) ^{2}\right\rangle  \\
&=&\frac{2}{\left( 2\pi \right) ^{2}}\int_{-\infty }^{\infty }\frac{\hbar
\omega ^{3}}{2\varepsilon _{0}c^{5}}\coth \left( \frac{\hbar \omega }{2T}
\right) d\omega .
\end{eqnarray*}

This results in a contribution to the energy density of value
\begin{equation*}
-\frac{1}{2\mu _{0}}\left\langle S^{2}\right\rangle =-\frac{1}{\left( 2\pi
\right) ^{2}}\int_{-\infty }^{\infty }\frac{\hbar \omega ^{3}}{2c^{3}}\coth
\left( \frac{\hbar \omega }{2T}\right) d\omega .
\end{equation*}

Furthermore, for the electric field
\begin{equation*}
\mathbf{E}\left( \mathbf{r},t\right) =-\nabla \phi \left( \mathbf{r}
,t\right) -\frac{\partial \mathbf{A}\left( \mathbf{r},t\right) }{\partial t},
\end{equation*}
we need to evaluate
\begin{equation*}
\left\langle \left\vert \mathbf{E}\right\vert ^{2}\right\rangle
=\left\langle \nabla \phi \cdot \nabla \phi \right\rangle +\left\langle 
\frac{\partial \mathbf{A}}{\partial t}\cdot \frac{\partial \mathbf{A}}{
\partial t}\right\rangle .
\end{equation*}

We thus have 
\begin{eqnarray*}
\left\langle \nabla \phi \cdot \nabla \phi \right\rangle  &=&\lim_{\mathbf{r}
_{b}\rightarrow \mathbf{r}_{a}}\frac{1}{\left( 2\pi \right) ^{2}}
\int_{-\infty }^{\infty }\int_{-\infty }^{\infty }\frac{\partial }{\partial
r_{\beta }^{a}}\frac{\partial }{\partial r_{\beta }^{b}}\left\langle \phi
\left( \mathbf{r}_{a},\omega \right) \phi \left( \mathbf{r}_{b},\omega
^{\prime }\right) \right\rangle  \\
&&\times \exp \left[ -\left( \omega +\omega ^{\prime }\right) t\right]
d\omega d\omega ^{\prime } \\
&=&\frac{1}{\left( 2\pi \right) ^{2}}\int_{-\infty }^{\infty }\frac{\hbar
\omega ^{3}}{2\varepsilon _{0}c^{3}}\coth \left( \frac{\hbar \omega }{2T}
\right) d\omega ,
\end{eqnarray*}
and
\begin{eqnarray*}
\left\langle \frac{\partial \mathbf{A}}{\partial t}\cdot \frac{\partial 
\mathbf{A}}{\partial t}\right\rangle  &=&-\lim_{\mathbf{r}_{b}\rightarrow 
\mathbf{r}_{a}}\frac{1}{\left( 2\pi \right) ^{2}}\int_{-\infty }^{\infty
}\int_{-\infty }^{\infty }\left\langle A_{\beta }\left( \mathbf{r}
_{a},\omega \right) A_{\beta }\left( \mathbf{r}_{b},\omega ^{\prime }\right)
\right\rangle  \\
&&\times \omega \omega ^{\prime }\exp \left[ -\left( \omega +\omega ^{\prime
}\right) t\right] d\omega d\omega ^{\prime } \\
&=&\frac{3}{\left( 2\pi \right) ^{2}}\int_{-\infty }^{\infty }\frac{\hbar
\omega ^{3}}{2\varepsilon _{0}c^{3}}\coth \left( \frac{\hbar \omega }{2T}
\right) d\omega ,
\end{eqnarray*}
where summation over repeated vector indices is assumed from now on.
We thus obtain the contribution of the electric field to the energy density
\begin{equation*}
\frac{\varepsilon _{0}}{2}\left\langle \left\vert \mathbf{E}\right\vert
^{2}\right\rangle =\frac{1}{\left( 2\pi \right) ^{2}}\int_{-\infty }^{\infty
}\frac{\hbar \omega ^{3}}{c^{3}}\coth \left( \frac{\hbar \omega }{2T}\right)
d\omega .
\end{equation*}

For the magnetic field we need to evaluate
\begin{eqnarray*}
\left\langle \left\vert \mathbf{B}\right\vert ^{2}\right\rangle 
&=&\left\langle \left( \nabla \times \mathbf{A}\right) \cdot \left( \nabla
\times \mathbf{A}\right) \right\rangle  \\
&=&\left\langle \frac{\partial A_{\gamma }}{\partial r_{\beta }}\frac{
\partial A_{\gamma }}{\partial r_{\beta }}-\frac{\partial A_{\beta }}{
\partial r_{\gamma }}\frac{\partial A_{\gamma }}{\partial r_{\beta }}
\right\rangle .
\end{eqnarray*}

So we have
\begin{eqnarray*}
\left\langle \frac{\partial A_{\gamma }}{\partial r_{\beta }}\frac{\partial
A_{\gamma }}{\partial r_{\beta }}\right\rangle  &=&\lim_{\mathbf{r}
_{b}\rightarrow \mathbf{r}_{a}}\frac{1}{\left( 2\pi \right) ^{2}}
\int_{-\infty }^{\infty }\int_{-\infty }^{\infty }\frac{\partial }{\partial
r_{\beta }^{a}}\frac{\partial }{\partial r_{\beta }^{b}}\left\langle
A_{\gamma }\left( \mathbf{r}_{a},\omega \right) A_{\gamma }\left( \mathbf{r}
_{b},\omega ^{\prime }\right) \right\rangle  \\
&&\times \exp \left[ -\left( \omega +\omega ^{\prime }\right) t\right]
d\omega d\omega ^{\prime } \\
&=&\frac{3}{\left( 2\pi \right) ^{2}}\int_{-\infty }^{\infty }\frac{\hbar
\omega ^{3}}{2\varepsilon _{0}c^{5}}\coth \left( \frac{\hbar \omega }{2T}
\right) d\omega ,
\end{eqnarray*}
and 
\begin{eqnarray*}
\left\langle \frac{\partial A_{\beta }}{\partial r_{\gamma }}\frac{\partial
A_{\gamma }}{\partial r_{\beta }}\right\rangle  &=&\lim_{\mathbf{r}
_{b}\rightarrow \mathbf{r}_{a}}\frac{1}{\left( 2\pi \right) ^{2}}
\int_{-\infty }^{\infty }\int_{-\infty }^{\infty }\frac{\partial }{\partial
r_{\gamma }^{a}}\frac{\partial }{\partial r_{\beta }^{b}}\left\langle
A_{\beta }\left( \mathbf{r}_{a},\omega \right) A_{\gamma }\left( \mathbf{r}
_{b},\omega ^{\prime }\right) \right\rangle  \\
&&\times \exp \left[ -\left( \omega +\omega ^{\prime }\right) t\right]
d\omega d\omega ^{\prime } \\
&=&\frac{1}{\left( 2\pi \right) ^{2}}\int_{-\infty }^{\infty }\frac{\hbar
\omega ^{3}}{2\varepsilon _{0}c^{5}}\coth \left( \frac{\hbar \omega }{2T}
\right) d\omega .
\end{eqnarray*}

With this we obtain the contribution from the magnetic field to the energy density to be
\begin{equation*}
\frac{1}{2\mu _{0}}\left\langle \left\vert \mathbf{B}\right\vert
^{2}\right\rangle =\frac{1}{\left( 2\pi \right) ^{2}}\int_{-\infty }^{\infty
}\frac{\hbar \omega ^{3}}{2c^{3}}\coth \left( \frac{\hbar \omega }{2T}
\right) d\omega .
\end{equation*}

Note the peculiar result that this contribution is half of that of the
electric field, and has the same magnitude as that of the scalar field.
Besides, the contribution from the electric field is twice that evaluated
using Maxwell electrodynamics.

The total energy density is finally
\begin{eqnarray*}
\left\langle u\right\rangle  &=&\frac{1}{\left( 2\pi \right) ^{2}}
\int_{-\infty }^{\infty }\frac{\hbar \omega ^{3}}{c^{3}}\coth \left( \frac{
\hbar \omega }{2T}\right) d\omega  \\
&=&\frac{2}{\left( 2\pi \right) ^{2}}\int_{0}^{\infty }\frac{\hbar \omega
^{3}}{c^{3}}\coth \left( \frac{\hbar \omega }{2T}\right) d\omega ,
\end{eqnarray*}
which corresponds to the spectral energy density 
\begin{equation}
\frac{d\left\langle u\right\rangle }{d\omega }=\frac{\hbar \omega ^{3}}{2\pi
^{2}c^{3}}\coth \left( \frac{\hbar \omega }{2T}\right) =\left( \frac{\hbar
\omega }{2}+\frac{\hbar \omega }{e^{\hbar \omega /T}-1}\right) \frac{\omega
^{2}}{\pi ^{2}c^{3}}, \label{spectrum}
\end{equation}
which is precisely Planck's formula, plus the contribution from the zero-point energy.


\color{black}
Thus, the additional scalar degree of freedom does not add an independent positive Planck contribution to the spectrum. The ABE energy is not a sum of separately positive electric, magnetic and scalar oscillator energies: the enhanced electric contribution is exactly compensated by the negative scalar contribution, leaving the total Planck spectrum unchanged.
\color{black}

\section{Comparison with Maxwell electrodynamics}
\label{fdt_maxwell}

In this section we consider a system in which strict local conservation of charge applies, and is thus described by Maxwell electrodynamics.

To begin, we note that most of the results obtained in the last section apply as well to a system described by Maxwell equations in the Lorentz gauge. This is easily seen by noting that one can always add a ``zero'' term like $-\left(
\partial \rho /\partial t+\nabla \cdot \mathbf{j}\right) \phi $ to the
specific power $\mathbf{j}\cdot \mathbf{E}$ in the energy equation in
Maxwell electrodynamics, \color{black} so that equation (\ref{dEdtfin}) is still valid, \color{black}and that equations (\ref{potwaves}) are also valid in Maxwell theory
when the Lorentz gauge is used. This readily shows that the last two of equations (\ref{correlfull}) apply also in Maxwell electrodynamics.

However, due to the conservation of charge, fluctuations of charge and
current are not independent; equivalently, in the Lorenz-gauge representation used here,
fluctuations of the scalar and longitudinal vector potentials are constrained rather than independent, and one
degree of freedom is thus lost relative to ABE.

In fact, the Lorentz gauge used to derive relations (\ref{potwaves})
requires that
\begin{equation*}
\phi \left( \mathbf{r}_{a},\omega \right) =-\frac{ic^{2}}{\omega }\frac{
\partial }{\partial r_{\beta }^{a}}A_{\beta }\left( \mathbf{r}_{a},\omega
\right) ,
\end{equation*}
so that, for instance,
\begin{equation*}
\left\langle \phi \left( \mathbf{r}_{a},\omega \right) \phi \left( \mathbf{r}
_{b},\omega ^{\prime }\right) \right\rangle =-\frac{c^{4}}{\omega \omega
^{\prime }}\frac{\partial }{\partial r_{\beta }^{a}}\frac{\partial }{
\partial r_{\gamma }^{b}}\left\langle A_{\beta }\left( \mathbf{r}_{a},\omega
\right) A_{\gamma }\left( \mathbf{r}_{b},\omega ^{\prime }\right)
\right\rangle ,
\end{equation*}
which, using, the last of (\ref{correlfull}), is easily seen to give the same
self-correlation for the  scalar potential, second equation of (\ref{correlfull}).

However, the first of (\ref{correlfull}) is not valid now, but instead we have 
\begin{eqnarray*}
\left\langle \phi \left( \mathbf{r}_{a},\omega \right) A_{\gamma }\left( 
\mathbf{r}_{b},\omega ^{\prime }\right) \right\rangle  &=&-\frac{ic^{2}}{
\omega }\frac{\partial }{\partial r_{\beta }^{a}}\left\langle A_{\beta
}\left( \mathbf{r}_{a},\omega \right) A_{\gamma }\left( \mathbf{r}
_{b},\omega ^{\prime }\right) \right\rangle  \\
&=&-\frac{i}{\omega }\frac{\partial }{\partial r_{\beta }^{a}}F_{ab}\left(
\omega \right) \coth \left( \frac{\hbar \omega }{2T}\right) \delta \left(
\omega +\omega ^{\prime }\right) .
\end{eqnarray*}

We thus have
\begin{eqnarray*}
\left\langle \nabla \phi \cdot \frac{\partial \mathbf{A}}{\partial t}
\right\rangle  &=&-\lim_{\mathbf{r}_{b}\rightarrow \mathbf{r}_{a}}\frac{1}{
\left( 2\pi \right) ^{2}}\int_{-\infty }^{\infty }\int_{-\infty }^{\infty
}i\omega ^{\prime }\frac{\partial }{\partial r_{\gamma }^{a}}\left\langle
\phi \left( \mathbf{r}_{a},\omega \right) A_{\gamma }\left( \mathbf{r}
_{b},\omega ^{\prime }\right) \right\rangle  \\
&&\times \exp \left[ -\left( \omega +\omega ^{\prime }\right) t\right]
d\omega d\omega ^{\prime } \\
&=&-\frac{1}{\left( 2\pi \right) ^{2}}\int_{-\infty }^{\infty }\frac{\hbar
\omega ^{3}}{2\varepsilon _{0}c^{5}}\coth \left( \frac{\hbar \omega }{2T}
\right) d\omega .
\end{eqnarray*}

In this way,
\begin{equation*}
\left\langle \left\vert \mathbf{E}\right\vert ^{2}\right\rangle
=\left\langle \nabla \phi \cdot \nabla \phi \right\rangle +\left\langle 
\frac{\partial \mathbf{A}}{\partial t}\cdot \frac{\partial \mathbf{A}}{
\partial t}\right\rangle +2\left\langle \nabla \phi \cdot \frac{\partial 
\mathbf{A}}{\partial t}\right\rangle ,
\end{equation*}
and the contribution of the electric field to the energy density is now
\begin{equation*}
\frac{\varepsilon _{0}}{2}\left\langle \left\vert \mathbf{E}\right\vert
^{2}\right\rangle =\frac{1}{\left( 2\pi \right) ^{2}}\int_{-\infty }^{\infty
}\frac{\hbar \omega ^{3}}{2c^{3}}\coth \left( \frac{\hbar \omega }{2T}
\right) d\omega ,
\end{equation*}
half of that obtained for ABE, and equal to the contribution of the magnetic
field, as it must in Maxwell electrodynamics, giving also the correct energy spectrum (\ref{spectrum}).

\section{Fluctuations in a conductor with non-conserved currents}
\label{fdt_conductor}

In this section we use a simple constitutive relation for the conserved and non-conserved sources in a non-polarizable, conducting medium. The model used, referred to as the ``$\gamma$-model", can be heuristically motivated by considering that, in addition to the current density $\mathbf{j}$ that accounts for the conventionally transported charge, a ``non-localized" current density $\mathbf{j}_{nl}$ quantifies the additional variation of charge density. Furthermore, for a homogeneous and isotropic medium the simplest possible constitutive relation between both currents is of the form $\mathbf{j}_{nl}=\gamma \mathbf{j}$, with $\gamma$ a constant scalar characteristic of the considered material medium. The local charge balance is thus described in this model by
\begin{equation}
 \frac{\partial\rho}{\partial t}+\left(1+\gamma\right)\nabla \cdot \mathbf{j}=0.  
 \label{gamma_model}
\end{equation}
A more formal derivation of this relation can be found in \cite{minotti_modanese_epjc2023}, while in \cite{minotti_modanese_math13050892} the same expression is obtained, but derived considering a possible UV dependence of Feynman graphs in the non-equilibrium Green's function formalism of fermion systems.

We start also here with the integrated expression of the conservation of energy in ABE: 
\begin{equation*}
\frac{dU}{dt}+\int \mathbf{S}_{u}\cdot d\mathbf{S}=-\int \left( \mathbf{j}
\cdot \mathbf{E}-I\phi \right) dV.
\end{equation*}

For a non-polarizable conductor, using the $\gamma $-model, we can write from (\ref{gamma_model}) 
\begin{equation*}
I=\frac{\partial\rho}{\partial t}+\nabla \cdot \mathbf{j}=-\gamma \nabla \cdot \mathbf{j},
\end{equation*}
resulting in
\begin{equation*}
-I\phi =\gamma \phi \nabla \cdot \mathbf{j}=\nabla \cdot \left( \gamma \phi 
\mathbf{j}\right) -\gamma \mathbf{j}\cdot \nabla \phi .
\end{equation*}
As before, since the volume of integration includes all sources, one has at
the surface $\mathbf{j}\cdot d\mathbf{S}=0$, leading to 
\begin{equation*}
\int -I\phi dV=\int -\gamma \mathbf{j}\cdot \nabla \phi dV=\int \gamma 
\mathbf{j}\cdot \left( \mathbf{E}+\frac{\partial \mathbf{A}}{\partial t}
\right) dV,
\end{equation*}
where the relation $\mathbf{E}=-\nabla \phi -\partial \mathbf{A}/
\partial t$, was used.

With all this, by formally defining as $\mathbf{F}$ the ``time integral" of $
\mathbf{E}$, so that $\mathbf{E}=\partial \mathbf{F}/\partial t$, we have 
\begin{equation}
\frac{dU}{dt}+\int \mathbf{S}_{u}\cdot d\mathbf{S}=-\int \mathbf{j}\cdot 
\frac{\partial }{\partial t}\left[ \left( 1+\gamma \right) \mathbf{F}+\gamma 
\mathbf{A}\right] dV.  \label{dEdtfin2}
\end{equation}

As in the previous section, for the application of the FDT, one can consider
a discrete set of small volumes $\triangle V$, at spatial positions $\mathbf{r}_{a}$, and identify the discrete variables 
\begin{eqnarray*}
x_{a}\left( t\right) &=&\text{ }j_{\beta }\left( \mathbf{r}_{a},t\right) , \\
f_{a}\left( t\right) &=&-\left[ \left( 1+\gamma \right) F_{\beta }\left( 
\mathbf{r}_{a},t\right) +\gamma A_{\beta }\left( \mathbf{r}_{a},t\right) 
\right] \triangle V.
\end{eqnarray*}

Further, the equation for the vector potential in the non-polarizable
conductor is also (\ref{Awave}), with general solution, in the frequency
domain, given by (\ref{Aaw}), which we rewrite here:
\begin{equation}
A_{\beta }\left( \mathbf{r}_{a},\omega \right) =\frac{\mu _{0}}{4\pi }\sum\limits_{b}\frac{
\exp \left( i\omega \left\vert \mathbf{r}_{a}-\mathbf{r}_{b}\right\vert
/c\right) }{\left\vert \mathbf{r}_{a}-\mathbf{r}_{b}\right\vert }j_{\beta
}\left( \mathbf{r}_{b},\omega \right) \triangle V.  \label{Abetaaw}
\end{equation}

For a conductor of constant, frequency-independent conductivity $\sigma $,
we have 
\begin{equation}
j_{\beta }\left( \mathbf{r}_{b},\omega \right) =-i\omega \sigma F_{\beta
}\left( \mathbf{r}_{b},\omega \right) ,  \label{Ohm}
\end{equation}
where it was used that, by definition, $\mathbf{E}\left( \mathbf{r}
_{b},\omega \right) =-i\omega \mathbf{F}\left( \mathbf{r}_{b},\omega \right) 
$.

We can thus write (\ref{Abetaaw}) as
\begin{equation*}
A_{\beta }\left( \mathbf{r}_{a},\omega \right) =-\frac{i\omega \mu
_{0}\sigma }{4\pi }\sum\limits_{b}\frac{\exp \left( i\omega \left\vert \mathbf{r}_{a}-
\mathbf{r}_{b}\right\vert /c\right) }{\left\vert \mathbf{r}_{a}-\mathbf{r}
_{b}\right\vert }F_{\beta }\left( \mathbf{r}_{b},\omega \right) \triangle V,
\end{equation*}
which allows us to write
\begin{equation*}
f_{a}\left( \omega \right) =-\sum\limits_{b}\left[ \left( 1+\gamma \right) \delta \left( 
\mathbf{r}_{a}-\mathbf{r}_{b}\right) -\frac{i\gamma \omega \mu _{0}\sigma }{
4\pi }\frac{\exp \left( i\omega \left\vert \mathbf{r}_{a}-\mathbf{r}
_{b}\right\vert /c\right) }{\left\vert \mathbf{r}_{a}-\mathbf{r}
_{b}\right\vert }\right] F_{\beta }\left( \mathbf{r}_{b},\omega \right)
\left( \triangle V\right) ^{2},
\end{equation*}
where it was used that $\delta _{ab}=\delta \left( \mathbf{r}_{a}-\mathbf{r}
_{b}\right) \triangle V$. On the other hand, with the same representation of 
$\delta _{ab}$, we can write (\ref{Ohm}) as 
\begin{equation}
j_{\beta }\left( \mathbf{r}_{a},\omega \right) =-i\omega \sigma \sum\limits_{b}\delta
\left( \mathbf{r}_{a}-\mathbf{r}_{b}\right) F_{\beta }\left( \mathbf{r}
_{b},\omega \right) \triangle V,  \label{Ohmv2}
\end{equation}
which indicates the convenience of writing
\begin{equation*}
f_{a}\left( \omega \right) =-\sum\limits_{b}\Lambda _{ab}F_{\beta }\left( \mathbf{r}
_{b},\omega \right) \triangle V,
\end{equation*}
with
\begin{equation}
\Lambda _{ab}=\left[ \left( 1+\gamma \right) \delta \left( \mathbf{r}_{a}-
\mathbf{r}_{b}\right) -\frac{i\gamma \omega \mu _{0}\sigma }{4\pi }\frac{
\exp \left( i\omega \left\vert \mathbf{r}_{a}-\mathbf{r}_{b}\right\vert
/c\right) }{\left\vert \mathbf{r}_{a}-\mathbf{r}_{b}\right\vert }\right]
\triangle V,  \label{Lambdaab}
\end{equation}
so that relation (\ref{Ohmv2}) can be written as
\begin{equation*}
j_{\beta }\left( \mathbf{r}_{a},\omega \right) =i\omega \sigma \sum\limits_{b,c}\delta \left( 
\mathbf{r}_{a}-\mathbf{r}_{b}\right) \Lambda _{bc}^{-1}f_{c}\left( \omega
\right) ,
\end{equation*}
where
\begin{equation}
\sum\limits_{c}\Lambda _{ac}^{-1}\Lambda _{cb}=\delta _{ab}=\delta \left( \mathbf{r}_{a}-
\mathbf{r}_{b}\right) \triangle V,  \label{Lm1L}
\end{equation}
which finally allows us to express the generalized susceptibility tensor as
\begin{equation*}
\alpha _{ab}\left( \omega \right) =i\omega \sigma \sum\limits_{c}\delta \left( \mathbf{r}
_{a}-\mathbf{r}_{c}\right) \Lambda _{cb}^{-1}.
\end{equation*}

If we write
\begin{equation*}
\Lambda _{cb}^{-1}=\frac{\delta \left( \mathbf{r}_{c}-\mathbf{r}_{b}\right) 
}{1+\gamma }+\Theta \left( \mathbf{r}_{c},\mathbf{r}_{b}\right) ,
\end{equation*}
with a function $\Theta $ to be determined, relations (\ref{Lambdaab}) and (\ref{Lm1L}) give
\begin{eqnarray*}
\frac{i\gamma \omega \mu _{0}\sigma }{4\pi \left( 1+\gamma \right) }\frac{
\exp \left( i\omega \left\vert \mathbf{r}_{c}-\mathbf{r}_{b}\right\vert
/c\right) }{\left\vert \mathbf{r}_{c}-\mathbf{r}_{b}\right\vert } &=&\left(
1+\gamma \right) \Theta \left( \mathbf{r}_{c},\mathbf{r}_{b}\right) \\
&&-\sum\limits_{d}\Theta \left( \mathbf{r}_{c},\mathbf{r}_{d}\right) \frac{i\gamma \omega
\mu _{0}\sigma }{4\pi }\frac{\exp \left( i\omega \left\vert \mathbf{r}_{d}-
\mathbf{r}_{b}\right\vert /c\right) }{\left\vert \mathbf{r}_{d}-\mathbf{r}
_{b}\right\vert },
\end{eqnarray*}
from which we can determine $\Theta $ iteratively in powers of the small dimensionless factor $\gamma $.

To first order in $\gamma $ we have 
\begin{equation*}
\Theta \left( \mathbf{r}_{c},\mathbf{r}_{b}\right) =\frac{i\gamma \omega \mu
_{0}\sigma }{4\pi \left( 1+\gamma \right) ^{2}}\frac{\exp \left( i\omega
\left\vert \mathbf{r}_{c}-\mathbf{r}_{b}\right\vert /c\right) }{\left\vert 
\mathbf{r}_{c}-\mathbf{r}_{b}\right\vert },
\end{equation*}
allowing us to write at this order
\begin{equation*}
\alpha _{ab}\left( \omega \right) =i\omega \sigma \left[ \frac{\delta \left( 
\mathbf{r}_{a}-\mathbf{r}_{b}\right) }{1+\gamma }+\frac{i\gamma \omega \mu
_{0}\sigma }{4\pi \left( 1+\gamma \right) ^{2}}\frac{\exp \left( i\omega
\left\vert \mathbf{r}_{a}-\mathbf{r}_{b}\right\vert /c\right) }{\left\vert 
\mathbf{r}_{a}-\mathbf{r}_{b}\right\vert }\right] .
\end{equation*}

We thus have, from (\ref{xaxb}),
\begin{equation}
\left( j_{\beta }^{\mathbf{r}_{a}}j_{\gamma }^{\mathbf{r}_{b}}\right)
_{\omega }=\sigma \hbar \omega \left[ \frac{\delta \left( \mathbf{r}_{a}-
\mathbf{r}_{b}\right) }{1+\gamma }-\frac{\gamma \sigma \mu _{0}\omega }{4\pi
\left( 1+\gamma \right) ^{2}}\frac{\sin \left( \omega \left\vert \mathbf{r}
_{a}-\mathbf{r}_{b}\right\vert /c\right) }{\left\vert \mathbf{r}_{a}-\mathbf{
r}_{b}\right\vert }\right] \coth \left( \frac{\hbar \omega }{2T}\right)
\delta _{\beta \gamma },  \label{jajbw}
\end{equation}
and, from (\ref{xawxbwp}),
\begin{eqnarray}
\left\langle j_{\beta }\left( \mathbf{r}_{a},\omega \right) j_{\gamma
}\left( \mathbf{r}_{b},\omega ^{\prime }\right) \right\rangle &=&2\pi \sigma
\hbar \omega \left[ \frac{\delta \left( \mathbf{r}_{a}-\mathbf{r}_{b}\right) 
}{1+\gamma }-\frac{\gamma \sigma \mu _{0}\omega }{4\pi \left( 1+\gamma
\right) ^{2}}\frac{\sin \left( \omega \left\vert \mathbf{r}_{a}-\mathbf{r}
_{b}\right\vert /c\right) }{\left\vert \mathbf{r}_{a}-\mathbf{r}
_{b}\right\vert }\right]  \notag \\
&&\times \coth \left( \frac{\hbar \omega }{2T}\right) \delta \left( \omega
+\omega ^{\prime }\right) \delta _{\beta \gamma }.  \label{avgjbjg}
\end{eqnarray}

These results reduce to the well know Nyquist relations for $\gamma =0$. The
interesting novelty is the addition of a non-local term that could in
principle be detected in experiments.

Writing the non-local term in the square brackets of (\ref{avgjbjg}) as 
\begin{equation*}
\frac{\gamma \sigma \mu _{0}\omega ^{2}}{4\pi c\left( 1+\gamma \right) ^{2}}
\frac{\sin \left( \omega \left\vert \mathbf{r}_{a}-\mathbf{r}_{b}\right\vert
/c\right) }{\omega \left\vert \mathbf{r}_{a}-\mathbf{r}_{b}\right\vert /c},
\end{equation*}
we see that for $\left\vert \mathbf{r}_{a}-\mathbf{r}_{b}\right\vert $ small
compared to $c/\omega $, which is of the order of cm for GHz frequencies,
this term is almost independent of the separation, and of the order of
magnitude
\begin{equation*}
\frac{\gamma \sigma \mu _{0}\omega ^{2}}{4\pi c}=\gamma \frac{\sigma \mu
_{0}c^{2}}{4\pi \omega }\left( \frac{\omega }{c}\right) ^{3},
\end{equation*}
which, taking into account that for a good conductor $\frac{\sigma \mu
_{0}c^{2}}{4\pi \omega }\gg 1$, could result in measurable effects, even for
small $\gamma $.

For instance, we consider the electrical noise detected at the terminals of
a cylindrical conductor of length $L$ and cross section $\Sigma $, in terms
of the spectral density of voltage fluctuations measured in open-circuit
conditions.  The voltage difference across a length $L$ is given as
\begin{equation*}
\triangle \mathcal{V}=-\int_{0}^{L}\frac{\mathbf{j}\cdot d\mathbf{l}}{\sigma 
},
\end{equation*}
so that we have
\begin{equation*}
\left\langle \triangle \mathcal{V}\left( \omega \right) \triangle \mathcal{V}
\left( \omega ^{\prime }\right) \right\rangle =\frac{1}{\sigma ^{2}\Sigma }
\int_{\Sigma }\int_{0}^{L}\int_{0}^{L}\left\langle j_{3}\left( \mathbf{r}
_{a},\omega \right) j_{3}\left( \mathbf{r}_{b},\omega ^{\prime }\right)
\right\rangle dz_{a}dz_{b}d\Sigma ,
\end{equation*}
where, in order to include all contributions from the bulk, we have taken
the average over the cross section of the conductor, and the line
integration is along the axis of the cylinder.

To simplify the derivations we consider low enough frequencies that satisfy $
\omega L/c\ll 1$ and $\omega \sqrt{\Sigma }/c\ll 1,$ so that 
\begin{equation*}
\frac{\sin \left( \omega \left\vert \mathbf{r}_{a}-\mathbf{r}_{b}\right\vert
/c\right) }{\omega \left\vert \mathbf{r}_{a}-\mathbf{r}_{b}\right\vert /c}
\simeq 1.
\end{equation*}
Using this, and the relation $dz_{b}d\Sigma =dV_{b}$, we directly obtain
from (\ref{avgjbjg})
\begin{equation*}
\left\langle \triangle \mathcal{V}\left( \omega \right) \triangle \mathcal{V}
\left( \omega ^{\prime }\right) \right\rangle =2\pi \hbar \omega \left[ 
\frac{L}{\left( 1+\gamma \right) \sigma \Sigma }-\frac{\gamma \mu _{0}\omega
^{2}L^{2}}{4\pi c\left( 1+\gamma \right) ^{2}}\right] \coth \left( \frac{
\hbar \omega }{2T}\right) \delta \left( \omega +\omega ^{\prime }\right) .
\end{equation*}

For not too low temperatures, and the low frequencies considered we can
approximate
\begin{equation*}
\hbar \omega \coth \left( \frac{\hbar \omega }{2T}\right) \simeq 2T,
\end{equation*}
and, since the conductor electric resistance is $R=L/\left( \sigma \Sigma
\right) $, we obtain 
\begin{equation}
\left\langle \triangle \mathcal{V}\left( \omega \right) \triangle \mathcal{V}
\left( \omega ^{\prime }\right) \right\rangle =4\pi T\left[ \frac{R}{\left(
1+\gamma \right) }-\frac{\gamma \mu _{0}\omega ^{2}L^{2}}{4\pi c\left(
1+\gamma \right) ^{2}}\right] \delta \left( \omega +\omega ^{\prime }\right)
.
\label{deltaVdeltaV}
\end{equation}

For $\gamma =0$ this is the classic, white noise Nyquist result. The interesting addition
is the violet noise term, proportional to\ $\omega ^{2}$, where the only
involved properties of the conductor are its length and its $\gamma $
parameter. For a conductor of characteristic size $L\simeq 1$ cm, and $
\omega \simeq 2\pi \times 10^{9}$ rad s$^{-1}$, we have
\begin{equation*}
\frac{\mu _{0}\omega ^{2}L^{2}}{4\pi c}\simeq 1.3\text{ }\Omega \text{.}
\end{equation*}

\section{Possible experimental test through thermal-noise spectroscopy}
\label{tests}

Thermal electrical noise has been experimentally investigated since the original measurements of Johnson, which established the existence of equilibrium voltage fluctuations in conductors, and the theoretical analysis of Nyquist, which related their spectral density to resistance and temperature. More recently, Johnson-noise thermometry has evolved into a high-precision metrological technique, with wide-band spectral analysis and two-channel cross-correlation methods allowing the extraction of extremely weak equilibrium noise signals with controlled uncertainties \cite{Benz2009JNT,Tew2019JNT}. In this sense, the possible experimental test suggested by the present work would not require a fundamentally new type of measurement, but rather the search for a specific and reproducible departure from the standard flat Nyquist spectrum in a conductor admitting a small degree of local charge non-conservation.

The result obtained in the previous section suggests a possible experimental signature of the $\gamma$-model in the form of a violet correction to the standard Johnson–Nyquist voltage-noise spectrum. In the low-frequency regime considered here, the open-circuit voltage fluctuations across a conductor of length $L$ are given by eq.\ \eqref{deltaVdeltaV}.

It is useful to rewrite the correction as an effective frequency-dependent contribution
\begin{equation}
\Delta R_\gamma(\omega)=
-\frac{\gamma}{(1+\gamma)^2}
\frac{\mu_0\omega^2L^2}{4\pi c},
\label{f-dependent}
\end{equation}
so that the relative deviation from the ordinary Nyquist spectrum is approximately
\begin{equation}
\eta(\omega)
\equiv
\frac{|\Delta R_\gamma(\omega)|}{R/(1+\gamma)}
\approx
\frac{\gamma}{1+\gamma}
\frac{\mu_0\omega^2L^2}{4\pi c R}.
\label{relative-deviation}
\end{equation}
This expression shows that the effect is enhanced by increasing frequency and sample length, and by decreasing the sample resistance, whereas the condition $\omega L/c\ll 1$ prevents one from increasing $L$ arbitrarily at fixed frequency.

For a representative sample length $L\simeq 1$ cm and frequency $f=\omega/2\pi\simeq 1$ GHz, one finds as we have seen
\begin{equation}
\frac{\mu_0\omega^2L^2}{4\pi c}\simeq 1.3\ \Omega.
\end{equation}
Accordingly, the relative corrections to the spectral density are of the same order of $\gamma$ for a sample with $R \sim 1 \Omega$.

These estimates suggest that the most favorable regime is that of relatively short, low-resistance conductors, probed in the spectral range from a few $10^8$ Hz up to about $10^9$ Hz. In this range the approximations $\omega L/c\ll 1$ and $\omega \sqrt{\Sigma}/c\ll 1$ may still remain plausible for centimeter-scale samples, while the $\omega^2$ dependence already provides a significant enhancement of the correction.

A natural experimental strategy would be to adapt the methods developed in precision Johnson-noise thermometry, in particular two-channel cross-correlation schemes with wide-band spectral analysis. Such techniques are specifically designed to suppress amplifier noise and to extract weak equilibrium noise spectra with high accuracy. In the present case one would not search for the absolute value of thermal noise itself, but rather for a reproducible departure from a flat spectrum.

In the classical regime considered in this paper, both the standard Nyquist term and the $\gamma$-dependent correction scale linearly with temperature. Therefore lowering the temperature does not significantly improve their relative ratio. Nevertheless, cryogenic operation may still be advantageous in practice, because it reduces the absolute noise level, facilitates low-noise amplification, and improves the overall control of the measurement.

\color{black}
A detailed experimental design lies beyond the scope of the present work. Still, the previous estimates indicate that values of $\gamma$ of order $10^{-3}$  could in principle become accessible in a dedicated experiment combining sub-ohm samples, centimeter-scale lengths, and spectral measurements approaching the GHz range. This should not be interpreted as an existing experimental bound, since conventional Johnson-noise thermometry generally operates in a different frequency, resistance, and geometrical regime.

    \subsection{Relation to existing experimental constraints}

It is useful to clarify the relation between the effects considered here and existing precision tests of electromagnetic fluctuations. The results obtained in Sec.\ \ref{fdt_ab} do not predict a modification of the total equilibrium black-body spectrum. As shown in Eq.\ \eqref{spectrum}, the total spectral energy density is exactly the Planck spectrum, including the zero-point contribution. The difference with respect to Maxwell electrodynamics concerns instead the partition of the total energy among the electric, magnetic, and Aharonov–Bohm scalar sectors. 
Precision measurements of the black-body spectrum constrain deviations of the total spectral density from the Planck distribution to a very high accuracy, of the order of a few parts in $10^5$
\cite{fixsen1996cosmic}. They do not, however, directly constrain a redistribution among the separate electric, magnetic, and scalar contributions when their sum remains unchanged.

Conventional black-body experiments use matter whose macroscopic charge and current densities satisfy LCC to very high accuracy. They therefore probe the Maxwell sector of ABE, for which $S=0$ under the usual boundary conditions and the standard field correlations are recovered.
Note that the result of Sec.\ \ref{fdt_ab} concerns instead the limiting theoretical case in which the fluctuating material sources themselves are unconstrained by LCC.

Similarly, most conventional tests involving antennas, microwave cavities, and cavity-QED systems employ sources that are locally conserved within their standard macroscopic description. For $I=\partial_t\rho+\boldsymbol{\nabla}\cdot\mathbf j=0$, Aharonov–Bohm electrodynamics reduces to the Maxwell theory relevant to such experiments. These measurements therefore do not directly test the material-dependent effect described by the $\gamma$-model, for which the scalar sector is excited only through a non-vanishing local source $I$. A dedicated analysis would be required to translate a particular cavity or antenna experiment into a bound on $\gamma$, since this parameter is not a universal constant of the theory but a property of the material system.

A more direct constraint can in principle be obtained from measurements of Johnson–Nyquist noise. For $|\gamma|\ll1$, Eq.\ \eqref{relative-deviation} gives
$$
\eta(\omega)
\simeq
\gamma\frac{\mu_0\omega^2L^2}{4\pi cR}.
$$
If a fit to the measured voltage-noise spectrum finds no quadratic-in-frequency contribution with a magnitude larger than a fraction $\delta$ of the ordinary Johnson–Nyquist spectral density at $\omega_{\max}$, the corresponding approximate bound is
$$
|\gamma|
\lesssim
\delta\frac{4\pi cR}{\mu_0\omega_{\max}^{2}L^{2}}.
$$
This expression also shows why existing high-accuracy Johnson-noise measurements do not necessarily provide a strong bound on $\gamma$. Such measurements are often optimized for thermometry and fundamental metrology, using relatively large sensing resistances and bandwidths in the sub-MHz range \cite{qu2015improved}. Since the predicted correction increases as $\omega^2L^2/R$, these conditions are not those that maximize the present effect.

For example, for $L=1 \ \mathrm{cm}$ and $f=1 \ \mathrm{GHz}$,
$$
\frac{\mu_0\omega^2L^2}{4\pi c}\simeq1.3\ \Omega.
$$
For a $1 \ \Omega$ sample, a measurement sensitive to a relative quadratic spectral correction of $10^{-3}$ would consequently probe values of $|\gamma|$ of order $8\times10^{-4}$. In contrast, at $f=550 \ \mathrm{kHz}$, for the same sample length, the corresponding coefficient is only approximately $4\times10^{-7} \ \Omega$. Even a parts-per-million measurement performed on a resistance of order $10^2\ \Omega$ would therefore not yield a competitive constraint on $|\gamma|\ll1$.

Existing measurements thus constrain very strongly ordinary departures from the Planck and Johnson–Nyquist laws, but they have generally not been optimized for the particular combination of frequency, sample length, resistance, and material dependence appearing in Eqs.\ \eqref{f-dependent} and \eqref{relative-deviation}. A meaningful experimental test would require comparing samples with different expected values of $\gamma$, while keeping their ordinary impedances and electromagnetic environments under control, and searching specifically for the predicted $L^2\omega^2$ scaling.
\color{black}

\section{Conclusions}
\label{conclusions}

Although the idea that charge could not be locally conserved may seem at first somehow startling, that possibility has been previously considered by many authors, notably by Aharonov and Bohm, and is now seriously discussed in the area of molecular devices. The main difficulty is that Maxwell electrodynamics is not applicable to systems with local non-conservation, so that an extended description that allows for it, and that coincides with Maxwell theory when charge is locally conserved, is required. Possibly the simplest extended theory that satisfies these requirements, along with that of relativistic covariance, is the electrodynamics of Aharonov-Bohm. We have thus used this particular theory to explore some of the observable consequences of a possible local non-conservation of charge.  

With the use of the Fluctuation Dissipation Theorem we have derived the energy spectrum of the electromagnetic field determined by the electrodynamics of Aharonov-Bohm for a system of material sources in thermal equilibrium, but which do not satisfy strict local conservation of charge (global charge is nonetheless conserved). We have also compared that derivation with the corresponding one for a system with strict local conservation of charge, described by Maxwell electrodynamics. A particularly interesting result is that both total energy spectra coincide, although the partition of the energy among the electric, magnetic, and scalar sectors differs.

In an actual situation local non-conservation of charge is expected to apply only to a small fraction of the material sources. The evaluation of the fluctuations for such a system thus requires a model for both kinds of sources and their interrelations. An example of this is presented for the case of a conductor in which the sources are described by a simple constitutive relation, for which we obtain the classical Johnson-Nyquist white noise result for voltage fluctuations, plus the addition of a small colored noise contribution. We further discuss the possibilities of an experimental test of this result.

A natural extension of the present analysis would be to electromagnetic theories characterized by a nonlocal gauge structure. In such theories the linear response is in general spatially nonlocal, so that the susceptibilities entering the fluctuation--dissipation relation should be replaced by response kernels depending on two spatial points, or equivalently on wave vector in translationally invariant systems. We do not expect spatial nonlocality by itself to modify the general structure of the fluctuation--dissipation theorem, provided that the theory admits a stable equilibrium state, a consistent Hamiltonian formulation, and microscopic reversibility. Rather, nonlocality would modify the relevant response functions and consequently the associated fluctuation spectra.   
The nonlocal gauge structure arises generally when a gauge theory with strict local interactions is extended to be nonlocal, either to enrich its estructure in order to solve some problems of the local theory \cite{partovi1982}, or as a procedure for its perturbative regularization \cite{evens1991}. While the extension to nonlocal interactions helps to regularize ultraviolet divergences, it generally ruins the original local gauge invariance (although not always \cite{chretien1954,partovi1982}), thus compromising the fundamental requirements of Poincaré invariance and unitarity. A further extension is thus required to make the theory nonlocal gauge-invariant, which results in a corresponding locally conserved current and the elimination of the ``nonphysical" degrees of freedom, like, for instance, photons with longitudinal polarization in nonlocal QED \cite{evens1991}. In this respect, a nonlocal gauge invariant theory would not present any ``anomalous" field polarization, so that the corresponding equilibrium spectrum, determined using the FDT, is not expected to show an anomaly as the one studied in this paper. Of course, this expectation has to be verified, and  a detailed application would therefore require specifying the particular nonlocal gauge theory and identifying its physical dynamical variables and conjugate generalized forces. This represents an interesting direction for future work.
\color{black}

\bibliography{fdt} 
\bibliographystyle{ieeetr}

\end{document}